 \documentclass[preprint2]{aastex}

\usepackage{graphicx}
\usepackage{multirow}
\usepackage{natbib}
\usepackage{color}
\usepackage{pdfpages}

\shorttitle{Two Mass Distributions in the L\,1641 MC}
\shortauthors{Polychroni, Schisano, Elia et al.}

\begin{document}


\title{Two Mass Distributions in the L\,1641 Molecular Clouds: The \emph{Herschel} connection of Dense Cores and Filaments in Orion A\thanks{\emph{Herschel} is an ESA space observatory with science instruments provided by European-led Principal Investigator consortia and with important pasrticipation from NASA.}}

\author{D. Polychroni\altaffilmark{1,3}, E. Schisano\altaffilmark{2,3}, D. Elia\altaffilmark{3}, A. Roy\altaffilmark{4}, S. Molinari\altaffilmark{3}, P. Martin\altaffilmark{5}, Ph. Andr\'e\altaffilmark{4}, D. Turrini\altaffilmark{3}, K.L.J. Rygl\altaffilmark{3}, J. Di Francesco\altaffilmark{6,7}, M. Benedettini\altaffilmark{3}, G. Busquet\altaffilmark{3}, A.M. di Giorgio\altaffilmark{3}, M. Pestalozzi\altaffilmark{3}, S. Pezzuto\altaffilmark{3}, D. Arzoumanian\altaffilmark{8}, S. Bontemps\altaffilmark{9}, M. Hennemann\altaffilmark{4,10}, T. Hill\altaffilmark{4}, V. K\"{o}nyves\altaffilmark{4,8}, A. Men'shchikov\altaffilmark{4}
F. Motte\altaffilmark{4}, Q. Nguyen-Luong\altaffilmark{5}, N. Peretto\altaffilmark{11}, N. Schneider\altaffilmark{9}, G. White\altaffilmark{12,13} }

\altaffiltext{1}{University of Athens, Department of Astrophysics, Astronomy and Mechanics, Faculty of Physics, Panepistimiopolis, 15784 Zografos, Athens, Greece.}
\altaffiltext{2}{Infrared Processing and Analysis Center, California Institute of Technology, Pasadena, CA, 91125, USA}
\altaffiltext{3}{Istituto di Astrofisica e Planetologia Spaziali (INAF-IAPS), via del Fosso del Cavaliere 100, 00133 Roma, Italy}
\altaffiltext{4}{Laboratoire AIM, CEA/IRFU CNRS/INSU Universit\'e Paris Diderot, Paris-Saclay, 91191 Gif-sur-Yvette, France}
\altaffiltext{5}{Canadian Institute for Theoretical Astrophysics, University of Toronto, 60 St. George Street, Toronto, ON M5S~3H8, Canada}
\altaffiltext{6}{National Research Council Canada, 5071 West Saanich Road, Victoria BC Canada, V9E 2E7} 
\altaffiltext{7}{Department of Physics \& Astronomy, University of Victoria, PO Box 355, STN, CSC, Victoria BC, V8W 3P6, Canada}
\altaffiltext{8}{IAS, CNRS (UMR 8617), Universit\'e Paris-Sud, B\^atiment 121, 91400 Orsay, France}
\altaffiltext{9}{Universit\'e de Bordeaux, Laboratoire d'\,Astrophysique de Bordeaux, CNRS/INSU, UMR 5804, BP 89, 33271, Floirac Cedex, France}
\altaffiltext{10}{MPIA Heidelberg, Germany}
\altaffiltext{11}{Cardiff School of Physics and Astronomy, Cardiff University, Queens Buildings, The Parade, Cardiff, Wales, CF24 3AA, UK}
\altaffiltext{12}{Space Science and Technology Department, Rutherford Appleton Laboratory, Chilton, Didcot, Oxon OX11 0QX, UK}
\altaffiltext{13}{Department of Physical Sciences, The Open University, Milton Keynes MK7 6AA, UK}

\email{dpolychroni@phys.uoa.gr}

\begin{abstract}
We present the \emph{Herschel} Gould Belt survey maps of the L\,1641 molecular clouds in Orion A. We extracted both the filaments and dense cores in the region. We identified which of dense sources are proto- or pre-stellar, and studied their association with the identified filaments. We find that although most (71\%) of the pre-stellar sources are located on filaments there is still a significant fraction of sources not associated with such structures. We find that these two populations (on and off the identified filaments) have distinctly different mass distributions. The mass distribution of the sources on the filaments is found to peak at 4\,$M_{\odot}$ and drives the shape of the CMF at higher masses, which we fit with a power law of the form d$N$/dlog$M \propto M^{-1.4\pm0.4}$. The mass distribution of the sources off the filaments, on the other hand, peaks at 0.8\,$M_{\odot}$ and leads to a flattening of the CMF at masses lower than $\sim$4\,$M_{\odot}$. We postulate that this difference between the mass distributions is due to the higher proportion of gas that is available in the filaments, rather than in the diffuse cloud.

\end{abstract}


\section{Introduction}

Results from recent large scale star formation surveys at far-infrared wavelengths (e.g. the Gould Belt survey, \citealp{andre10}; the Hi-GAL survey, \citealp{molinari10}) as well as surveys in molecular emission tracers (e.g. \citealp{falgarone98}) demonstrate the practically ubiquitous presence of filaments in star forming regions. Whether these are a consequence of turbulence (e.g. colliding flows; \citealp{klessen04}, \citealp{gong11}) or arise from instabilities in self-gravitating sheets, filaments are observed to have complex structures with branches and hubs \citep{myers09}. It is on such structures that most, but not all, dense cores are observed to exist (e.g. \citealp{sasha10}; \citealp{andre10}). Since dense cores are thought to be the precursors of proto-stars (e.g. \citealp{enoch06}) and that the core mass function (CMF) of such cores is similar in shape to the stellar initial mass function (IMF) (\citealp{motte98}; \citealp{motte01}; \citealp{alves07}; \citealp{konyves10}) it is imperative to understand how their immediate environment - here the presence of filaments - affects them and influences their mass distribution. 

In this Letter we continue the study of the Orion A molecular cloud (MC) complex, in the framework of the Herschel Gould Belt survey (HGBS) (\citealp{andre10}; \citealp{roy13}), focusing on the observations of the L\,1641 molecular clouds. Assuming the same distance as for the ONC of 414$\pm$7\,pc \citep{menten07}, L\,1641 is the southernmost part of the Orion A (MC), and constitutes the continuation of the integral shaped filament. With the exception of L\,1641--N it is considered to be a low to intermediate mass star formation region \citep{allen08}. Thus, L\,1641 is an ideal site to study the link between filaments and the formation of dense cores. Here, we report on the striking difference between the properties of cores located on filamentary structures compared to those in the rest of the cloud.

\section{Observations and Data Reduction}
As part of the Key Project ``HGBS", the L\,1641 region was imaged in October 2010 with the Herschel Space Observatory \citep{pilbratt10} in parallel mode, i.e. using PACS \citep{poglitsch10} at 70/160$\,\mu$m and SPIRE \citep{griffin10} at 250/350/500$\,\mu$m with a scanning speed of 60$\arcsec$/s, covering a common mapping area between PACS and SPIRE of 22.1 deg$^2$. Initial data reduction was performed using the Herschel Interactive Processing Environment (HIPE v. 7.0.0). The data were corrected for drift; identified glitches and flux discontinuities were also removed. The maps were then reconstructed using the Fortran code ROMAGAL (see \citet{traficante11} and  \citet{piazzo12} for details). The images were astrometrically registered on the MIPS 24\,$\mu$m images of the region \citep{megeath12} yielding an astrometric precision of $\sim$0.7$^{\arcsec}$. We have applied zero level offsets that were measured from Planck and IRAS data \citep{bernard10}.

\section{Analysis - The L\,1641 Molecular Clouds}

   \begin{figure*}[!ht]
   \centering
    \includegraphics[width=8.cm]{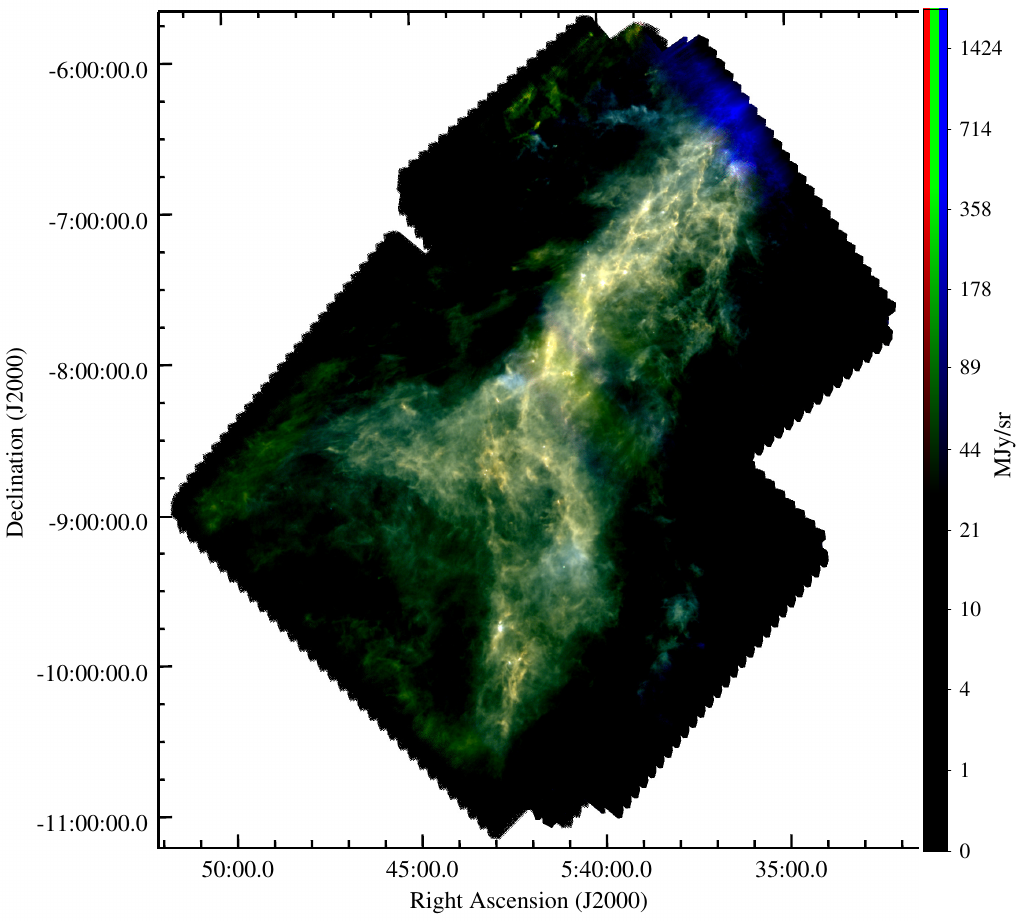}
    \includegraphics[width=8.cm]{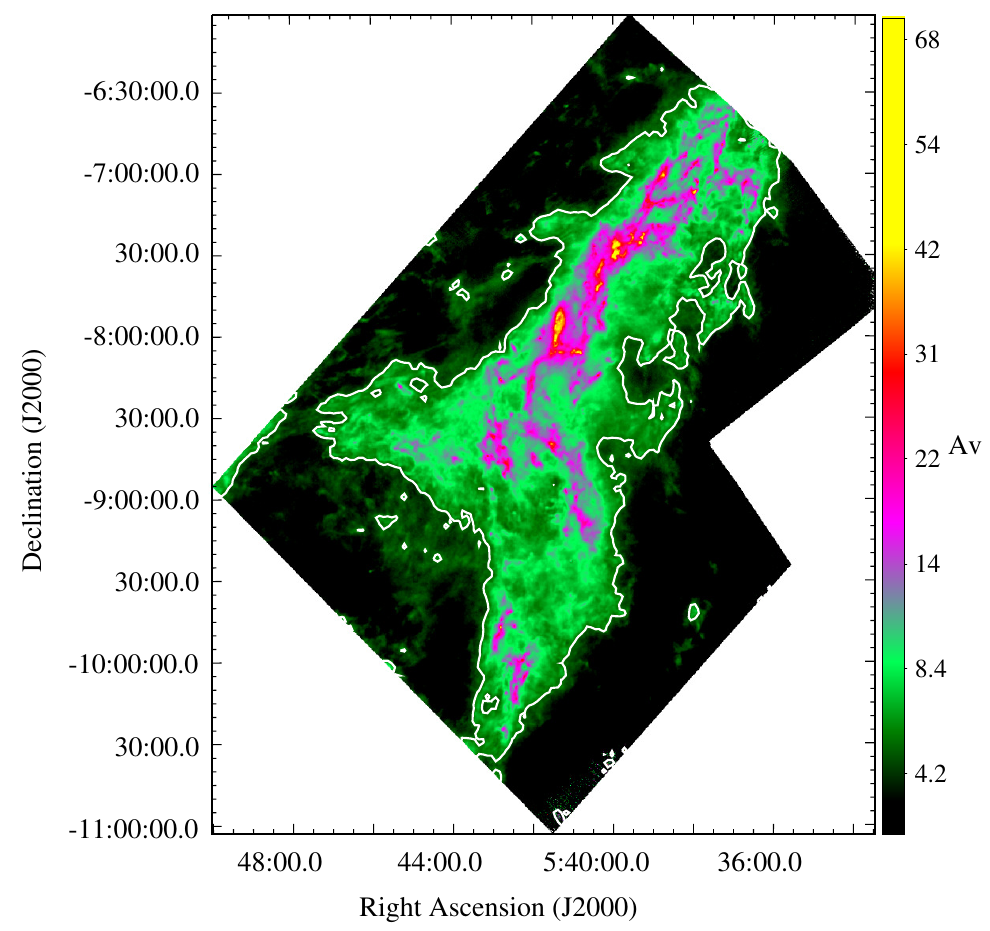}
    \caption{\textbf{Left:} The RGB (blue: 160\,$\mu$m; green: 250\,$\mu$m and red: 350\,$\mu$m) image of the L\,1641 MC. The intricate network of filaments that continue down from the integral shaped filament in the northern region of Orion A (not covered here) easily stands out. The scale is logarithmic to accommodate the wide intensity range of the map. \textbf{Right:} The column density map of the L\,1641 MCs expressed in units of extinction (the contour level is at 2\,A$_V$). The resolution of the map is 36$\arcsec$. A colour version of this image can be found in the online version of this letter.}
    \label{fig:mosaic}%
    \end{figure*}

In the left panel of Figure \ref{fig:mosaic} we present the RGB colour composite image of the L\,1641 clouds. We follow the \citet{allen08} nomenclature for these regions, where L\,1641 refers to all regions at Declinations below $\delta$(J2000)=--0.6$^{d}$10$^{m}$.  However, it should be noted that we have not included the L\,1641--N region. The \emph{Herschel} maps of the L\,1641 clouds show that it is permeated with a network of interconnecting filaments arranged in (partly) overlapping, V-shaped substructures. Most of the filaments lie on a coherent Northwest-to-Southeast axis, except in the southernmost parts of this map. Here, this generally collimated network of filaments splits in two directions, with some filamentary structures following an almost orthogonal direction to the above axis (i.e. from east to west). 

In the following analysis we use a dust opacity of $\mathrm{\kappa_{Thz}}$=0.1\,cm$^2$g$^{-1}$ (cf. \citealp{beckwith90}) and fix the dust emissivity, $\beta$, to 2 \citep{hildebrand83}, as is the standard adopted for the HGBS (e.g. \citealp{konyves10}), to derive the mass of the dense sources and the column density map of the region.

\subsection{The Filaments}
The filamentary shape of the dust distribution in the cloud is most evident in the column density map obtained by fitting greybody SEDs in each pixel of the Herschel 160 to 500\,$\mu$m maps (see the right panel of Figure \ref{fig:mosaic}). We identify the filamentary regions on the column density map by means of algorithms for pattern recognition. In short, such algorithms start from the second derivative of the map (see also \citealp{molinari10}), compute the eigenvalues of the Hessian matrix in each pixel and select the regions where the curvature along one of the eigendirections exceeds a certain threshold value. This threshold defines the minimum variation in the contrast that is accepted to separate a filamentary region from its surroundings. Afterwards, morphological operators are applied to determine the central pixels of the identified regions. A more detailed description of the method will be presented in Schisano et al. (submitted). Across the L\,1641 MCs the average deconvolved FWHM width of the filaments is 0.15\,pc, in agreement with \citet{arzoumanian11}. Their lengths vary from 0.5\,pc to $\sim$9\,pc. Temperatures along their ridge are found to be generally lower than the surrounding medium at 12--13\,K and the masses range from $\sim$5\,$M_{\sun}$ to 5$\times$10$^3$\,$M_{\sun}$. These values are in broad agreement with the findings of \citet{nagahama98}; any divergence is likely due to the lower resolution of their $^{13}$CO J=1$\to$0 data (2$\arcmin$ vs 36$\arcsec$) and the use of slightly different distances to the cloud (484\,pc instead of 414\,pc). 

\subsection{The Sources}
\label{sec:source}
We used the CUrvature Threshold EXtractor package (CuTEx; \citealp{molinari11}) to detect and extract, through Gaussian fitting and background subtraction, the sources individually from each band. We accept only those sources with a $S/N$ ratio higher than 5. We merge the five catalogues, following \citet{elia10}, associating sources across the bands if their positional distance is within the radius of the Herschel $HPBW$ at the longer wavelength. We do not consider cases of multiple associations, i.e. at shorter wavelengths we kept the source closest to the position of the centre of the equivalent source at the longer wavelength. We have not attempted to split the flux densities of the sources at longer wavelengths at such instances. We select only those sources that are found in three consecutive bands whose size at FWHM is less than 0.1\,pc or $\sim$\,50" across, in order to select single core sources rather than clumps.

In total we find 493 sources which we fit with a grey-body model to determine the masses and temperatures of these cores. The complete and definite catalogue of this region will be presented in a forthcoming publication.

\section{Results and Discussion}

\subsection{Column Density and Mass}

We derived the total mass of the area included within A${_V}$ of 2 magnitudes from the constructed column density map (Figure \ref{fig:mosaic}), where the conversion between column density and extinction is given by $N_{H_2}$=9.4$\times$10$^{20}$\,A$_V$ \citep{bohlin78}. We find a mass of 3.7$\times$10$^{4}$\,$M_{\sun}$ for the L\,1641 molecular clouds. Again, using the column density map we estimate the total mass within the identified filaments at 1.16$\times$10$^{4}$\,$M_{\sun}$ which represents 31.4\% of the total mass of the L\,1641 molecular clouds.

\subsection{Source Classification}
To identify the proto-stellar cores in our catalogue, we assume that objects with a 70\,$\mu$m detection are proto-stellar cores (\citealp{stutz13}; \citealp{dunham08}). Since the L\,1641 clouds have been observed by Spitzer in 24\,$\mu$m \citep{megeath12} we use these images and catalogue to determine whether our identified proto-stellar cores are also detected in 24\,$\mu$m. We find that all of our proto-stellar cores (109 objects) are also detected in 24\,$\mu$m, while none of the rest (384) have a 24\,$\mu$m counterpart. We, therefore, feel confident in our classification of objects with a 70\,$\mu$m detection as proto-stellar cores. 

We further differentiate between \emph{pre-stellar} cores (i.e. starless, gravitationally collapsing cores, e.g. \citealp{andre00}; \citealp{difrancesco07}), from the \emph{starless}, non-collapsing objects, potentially confined by external pressure, using the critical Bonnor-Ebert mass, $\mathrm{M_{BE}\approx\,2.4\,R_{BE}\,a^2/\,G}$ \citep{bonnor56}, where $a$ is the isothermal sound speed at the core temperature, given from the grey-body model fit, $G$ the gravitational constant, and R$_{BE}$ the Bonnor-Ebert radius. As R$_{BE}$ we use the deconvolved observed size of the sources, measured at 250\,$\mu$m. Following \citet{rygl13}, we define as pre-stellar cores those sources with $\mathrm{M_{obs}/M_{BE} \geq\,1}$. Setting this ratio to 0.5 as in \citet{konyves10} reclassifies 34 objects as pre-stellar, but does not alter the CMF nor our statistics in a significant way. In Table \ref{tab:prop} we summarise the physical properties of the selected objects across L\,1641. 

Finally, we split our cores depending on whether their position is within or outside a filament identified area. For simplicity we will henceforth call our cores \emph{on} or \emph{off} filament to distinguish between the two categories. We find that 67\% of our sample cores are located \emph{on} filaments, of which 229 are pre-stellar, 92 are starless and 83 are proto-stellar. Of the cores located \emph{off} filaments 19 are pre-stellar, 44 are starless and 26 are proto-stellar. In the following we will ignore the proto-stellar cores, unless specifically stating otherwise, and will concentrate only on the pre-stellar and starless cores.

\begin{table*}[!ht]
\label{tab:prop}
\renewcommand{\arraystretch}{1.2}
\centering
\caption{Summary of the properties of the pre-stellar and starless cores across the L\,1641 MCs. The quoted sizes are the deconvolved diameters of the cores measured at 250\,$\mu$m.}
\begin{tabular}{|l|cc|cc|c|}
\hline

& \multicolumn{2}{c}{Pre-stellar} & \multicolumn{2}{|c|}{Starless}& \\ \hline
Filament Location & ON &OFF &ON &OFF & \\  \hline \hline
Source Counts & 229 &92 &19 &44 &  \\ \hline
\multirow{2}{*}{Temperature (K)}&8.7& 8.8&13.2 &12.8 &mean\\
                                        &8.5& 8.7&12.9 &12.7 &median\\\hline
\multirow{2}{*}{Mass ($M_{\sun}$)} &6.3 &1.6 &0.3 &0.2& mean\\
                   &4.7 &1.4 &0.3 &0.2& median \\\hline
\multirow{2}{*}{Size (arcsec)} & 24.7&23.7 &26.5 &23.8& mean\\
               & 24.4&24.0 &25.1&23.8&median \\\hline
\end{tabular}
\end{table*}

\begin{figure}[!ht]
   \centering
   \includegraphics[width=7.5cm]{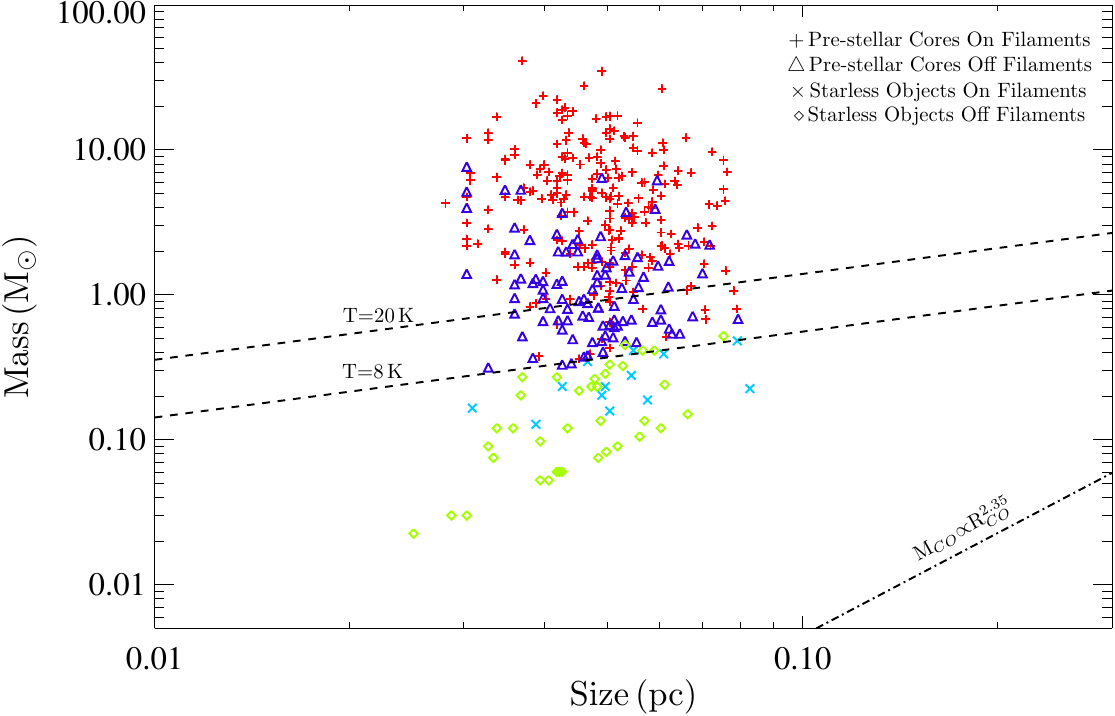}
   \caption{Mass versus size plot for the starless sources in the L\,1641 clouds. The dashed lines signify the Bonnor-Ebert mass for sources at 8\,K and 20\,K of different sizes. The dot-dash line signifies the Larson mass-size relation of CO clumps.A colour version of this image can be found in the online version of this letter.}
    \label{fig:ms}%
    \end{figure}

\begin{figure}[!ht]
   \centering
   \includegraphics[width=8.cm]{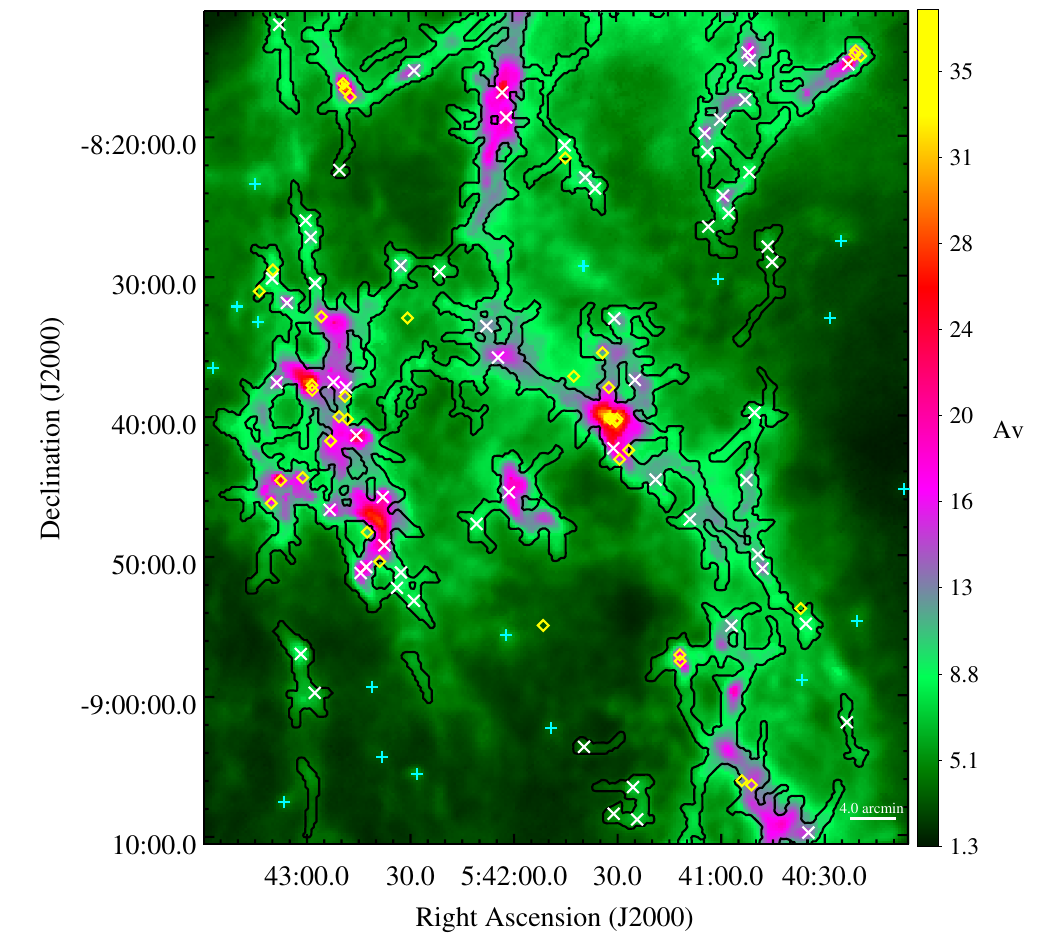}
   \caption{Close-up of a part of L\,1641 show in Fig 1. The black contours trace the filaments while the crosses denote the pre-stellar cores \emph{on} the filaments, the plus signs those \emph{off} them, and the diamonds the proto-stars. A colour version of this image can be found in the online version of this letter.}
    \label{fig:zoom}%
    \end{figure}

We find that 92\% of our sources \emph{on} filaments are pre-stellar, which drops to 68\% when considering sources \emph{off} filaments. Clearly, this is heavily dependent on the source environment, as well as on the distance as shown by \citet{elia13}. In Figure \ref{fig:zoom} we show an example of pre-stellar cores located \emph{on} filaments, as compared to those \emph{off} them. Note that all filaments with associated sources have A${_V}$\,$>$\,5 and most exhibit regions with extinctions in excess of 20 magnitudes. In Figure \ref{fig:ms} we plot the pre-stellar and starless cores in a mass vs. size diagram. We also plot the Larson relation between the mass and the size of a core given from CO luminosities \citep{elmegreen96} below which sources are supported by turbulent motions. All cores classified as pre-stellar are found to lie well above this line, indicative of the gravitational contraction these sources are presumably undergoing.

\subsection{Mass Distributions and Core Formation Efficiency}

\begin{figure*}[!ht]
   \centering
   \includegraphics[width=15cm]{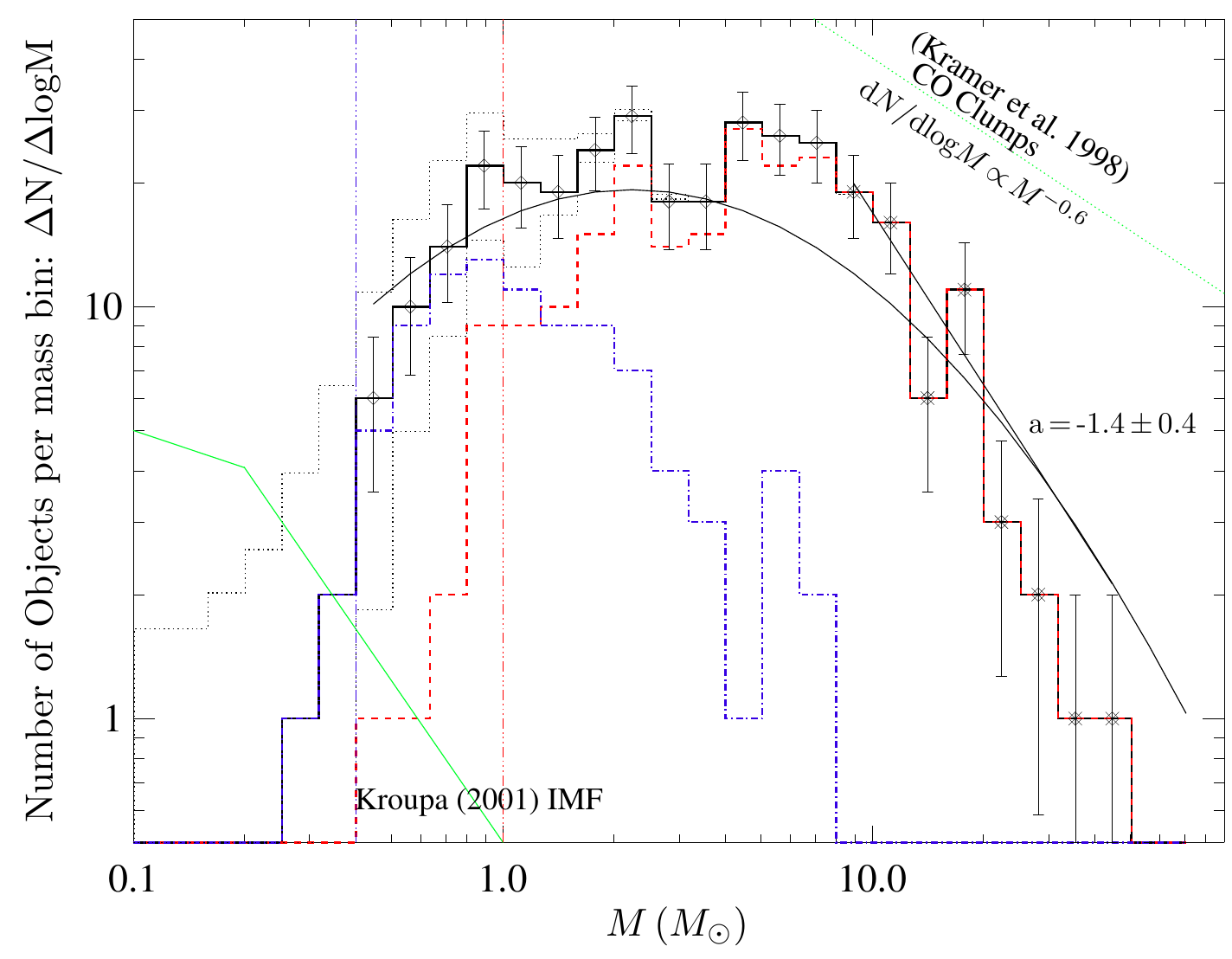}
    \caption{In black we plot the core mass function of the L\,1641 clouds. In red (dashed line) and blue (dash-dot line) are the mass distributions of the pre-stellar sources located \emph{on} and \emph{off} the identified filament regions respectively. The two vertical dashed lines represent the completeness limits of the field (at 0.4\,$M_{\sun}$) and of the filaments (at 1.0\,$M_{\sun}$). The black curve is the lognormal fit to the data (peaking at 2.2$\pm{0.05}$\,$M_{\sun}$ and with standard deviation of 0.6$\pm{0.05}$) while a linear fit to the high-mass end of the CMF gives a slope of -1.4$\pm$0.4. The green line indicates the \citet{kroupa01} initial mass function power law. A colour version of this image can be found in the online version of this letter.}
    \label{fig:cmffit}%
    \end{figure*}

In Figure \ref{fig:cmffit} we plot the mass distribution of the pre-stellar sources in L\,1641. The mass ranges between 0.2 and 55\,$M_{\sun}$. The distribution of masses is seen to be quite flat between 1\,$M_{\sun}$ and 4\,$M_{\sun}$ and cannot be fit by a simple power law. This observed break of the CMF at $M$\,$\sim4\,M_{\sun}$ agrees very well with the recent models of \citet{padoan11}. A power law fit to the high-mass end of the observed CMF (of the form dN/dlogM) gives a slope of -1.4$\pm$0.4, which is consistent within the errors with previous estimates for the Orion A MC (slope: -1.3; \citealp{ikeda07}). 

In the same Figure we plot the mass distributions of those pre-stellar cores \emph{on} filaments (71\%) and those \emph{off} them (29\%). The two distributions are found to peak at two different masses; 0.8\,$M_{\sun}$ for the objects located \emph{off} filaments and 4\,$M_{\sun}$ for those located \emph{on} the filaments. We find that the slope of the CMF at larger masses is driven by the sources located \emph{on} the filaments, while the flattening of the mass distribution at masses lower than 4\,$M_{\sun}$ is the result of sources not associated with filaments. An independent source extraction and photometry estimation in the field, using an alternative algorithm, $getsources$ \citep{sasha12}, yields the same behaviour for the masses of the pre-stellar sources \emph{on} and \emph{off} the filaments.  

The significance of this result is given by the completeness limit which we calculate by adding 900 synthetic sources blue characterised by the same size, mass and temperature distribution as the observed CMF in each of the maps. We extracted them using the same parameters as for the source extraction (section \ref{sec:source}). As there is a significant difference of column densities between the filaments and the rest of the L\,1641, we have calculated two mass completeness limits, derived from the 250\,$\mu$m flux limits. For the filaments we find that the completeness limit (at the 80\% level) is at 1.0\,$M_{\sun}$  while \emph{off} the filaments we are complete down to 0.4\,$M_{\sun}$. Therefore, the behaviour of the mass distributions of the two classes of cores should not be a result of incomplete sampling of cores. A Kolmogorov-Smirnov test confirms that the two distributions, above their respective completeness limits, are different with a certainty of 99.99\% (where the maximum deviation of the datasets is d=0.518, and the probability that they are drawn from the same distribution is p=3.7$\times$10$^{-12}$). 

To determine to which extent filaments contribute flux to the sources, we compared the input fluxes of the synthetic sources with the respective ones measured by CuTEx, using the SPIRE 250\,$\mu$m band as reference. We convert the input and measured fluxes to masses and bin them in the same way as the observed sources. For each bin we compute the fraction of synthetic sources within 20\% of the input value and from that we derive the 1--$\sigma$ uncertainty (due to background contribution) associated to each of the bin centres. 

We created 10$^5$ synthetic populations of 500 sources for the \emph{on} and \emph{off} filament mass distributions using Monte Carlo extractions. For each synthetic source we first determine to mass bin it belongs to. We then extract its measured mass value, assuming a gaussian distribution with the $\sigma$ parameter determined above and centred on the mass value of the bin centre. We rebin the mass distribution of each population as for the observed sources and for each bin we record the minimum and maximum values among the 10$^5$ populations for the \emph{on} and \emph{off} filament mass distribution and for the total one. Figure \ref{fig:cmffit} shows that the contribution from the background is relevant only for sources near the completeness limit or fainter. For these sources the 1--$\sigma$ relative error is of the level of 50\% or higher. Above the completeness limit, however, the background flux contribution is not significant enough to affect the results of this study.

Finally, we have measured the total mass within the pre-stellar cores located \emph{on} and \emph{off} the filaments. We find that, in total, 1.6$\times$10$^3$\,$M_{\sun}$ is held by such objects. Using the standard core formation efficiency (CFE) equation, $\mathrm{M_{cores}/(M_{cloud}+M_{cores})}$, where $\mathrm{M_{cloud}=}$ $\mathrm{3.7\times10^{4}\,M_{\sun}}$, we calculate that the CFE of the L\,1641 molecular clouds is 4\%, comparable with the results of \citet{elia13}. This value increases to 12\% when we consider only the dense cores \emph{on} filaments as well as the total mass within these filaments (i.e. where $\mathrm{M_{cores}=M_{on}}$ and $\mathrm{M_{cloud}=M_{filaments}}$). Although we do not derive masses for the proto-stars, since they are also preferentially located \emph{on} filaments (79\%), we can extrapolate that there will be increased star formation efficiency on filaments. This would agree well with the results of \citet{kryukova12} and \citet{gutermuth11} who find a higher SFE in higher density regions.

An interesting pattern seems to emerge concerning the properties of dense pre-stellar cores depending on their location \emph{on} or \emph{off} filaments.  The greater incidence of more gravitationally bound cores \emph{on} filaments can be explained by two effects.  First, fainter objects, i.e., ones of lower mass and hence more likely to be unbound, are not easy to detect towards filaments, as is demonstrated by the higher completeness limit in such structures. Second, cores located \emph{on} filaments find themselves in environments with larger reservoirs of mass, and possibly larger external pressures, than those \emph{off} filaments.  Indeed, though dense cores may form in the same general way \emph{on} or \emph{off} filaments, access to greater amounts of gas in filaments may be the root cause of the two separate core mass distributions noted in Figure \ref{fig:cmffit}. Furthermore, the denser background environment likely also results in the higher core formation efficiencies seen towards filaments, making them the preferred, but not unique, star formation site.

\section{Conclusions}

We have presented here the L\,1641 maps from the Herschel Gould Belt survey. These clouds have ``v-shaped'' filamentary structures that mostly follows a Northwest-to-Southeast axis similar to that of the integral shaped filament located in the north of these maps. While most (84\%) of the gravitationally bound pre-stellar sources are located on the identified filaments, star formation is also taking place in the more diffuse molecular cloud. We obtained the CMF of the region for which we found a slope of 1.4$\pm$0.4. We find that there are two distinct mass distributions that contribute to the CMF. The first is due to the dense sources \emph{on} the filaments, peaking at 4\,$M_{\sun}$, and drives the slope of the CMF in the higher mass end. The second peak (at 0.8\,$M_{\sun}$) is due to the dense sources \emph{off} the filaments and is responsible for the flattening of the CMF at masses lower than $\sim$4\,$M_{\sun}$. 

Furthermore we measured the CFE for all the L\,1641 molecular clouds at 4\% and we see it increases significantly (12\%) when considering only the filament identified regions and the pre-stellar objects located there. We conclude that these two mass distributions are the direct result of there being a larger reservoir of mass \emph{on} the filaments for the sources to accrete from, making star formation \emph{on} the filaments dominant but not exclusive.


%
\acknowledgments
We thank Dr. Despoina Hatzidimitriou and Scige John Liu for useful discussions and comments. We wish to thank the anonymous reviewer for comments that improved the manuscript.

PACS has been developed by a consortium of institutes led by MPE (Germany) and including UVIE (Austria); KU Leuven, CSL, IMEC (Belgium); CEA, LAM (France); MPIA (Germany); INAFIFSI /OAA/OAP/OAT, LENS, SISSA (Italy); IAC (Spain). This development has been supported by the funding agencies BMVIT (Austria), ESA-PRODEX (Belgium), CEA/CNES (France), DLR (Germany), ASI/INAF (Italy), and CICYT/MCYT (Spain). SPIRE has been developed by a consortium of institutes led by Cardiff University (UK) and including Univ. Lethbridge (Canada); NAOC (China); CEA, LAM (France); IFSI, Univ. Padua (Italy); IAC (Spain); Stockholm Observatory (Sweden); Imperial College London, RAL, UCLMSSL, UKATC, Univ. Sussex (UK); and Caltech, JPL, NHSC, Univ. Colorado (USA). This development has been supported by national funding agencies: CSA (Canada); NAOC (China); CEA, CNES, CNRS (France); ASI (Italy); MCINN (Spain); SNSB (Sweden); STFC (UK); and NASA (USA). HIPE is a joint development by the Herschel Science Ground Segment Consortium, consisting of ESA, the NASA Herschel Science Center, and the HIFI, PACS and SPIRE consortia. This research has made use of NASA's Astrophysics Data System.

DP is funded through the Operational Program ``Education and Lifelong Learning'' that is co-financed by the European Union (European Social Fund) and Greek national funds; KLJR, GB, DE and MP are funded by ASI fellowships under contract numbers I/005/11/00, I/038/08/0 and I/029/12/0.

\end{document}